\documentclass[conference]{IEEEtran}
\IEEEoverridecommandlockouts
\usepackage{cite}
\usepackage{upgreek}
\usepackage{amsmath,amssymb,amsfonts}
\usepackage{algorithmic}
\usepackage{graphicx}
\usepackage{textcomp}
\usepackage{url}
\usepackage{xcolor}
\usepackage{adjustbox}
\def\BibTeX{{\rm B\kern-.05em{\sc i\kern-.025em b}\kern-.08em
    T\kern-.1667em\lower.7ex\hbox{E}\kern-.125emX}}
\begin{document}

\title{Evaluating Driver Perceptions of Integrated Safety Monitoring Systems for Alcohol Impairment and Distraction\\
}

\author{\IEEEauthorblockN{RoshikNagaSai Patibandla}
\IEEEauthorblockA{\textit{MOT Charter School} \\
roshik.patibandla@gmail.com}
\and
\IEEEauthorblockN{Ross Greer}
\IEEEauthorblockA{\textit{University of California, Merced} \\
rossgreer@ucmerced.edu}
}

\maketitle
    
\begin{abstract}
The increasing number of accidents caused by alcohol-impaired driving has prompted the development of integrated safety systems in vehicles to monitor driver behavior and prevent crashes. This paper explores how drivers perceive these systems, focusing on their comfort, trust, privacy concerns, and willingness to adopt the technology. Through a survey of 115 U.S. participants, the study reveals a preference for non-intrusive systems, such as those monitoring eye movements, over more restrictive technologies like alcohol detection devices. Privacy emerged as a major concern, with many participants preferring local data processing and anonymity. Trust in these systems was crucial for acceptance, as drivers are more likely to adapt their behavior when they believe the system is accurate and reliable. To encourage adoption, it is important to address concerns about privacy and balance the benefits of safety with personal freedom. By improving transparency, ensuring reliability, and increasing public awareness, these systems could play a significant role in reducing road accidents and improving safety. 
\end{abstract}


\section{Introduction}

In the United States in 2020, alcohol-impaired driving caused one death every 45 minutes\cite{NHTSA2020}.  Drunk driving is a significant road safety issue worldwide. Traffic crashes involving one or more alcohol-impaired drivers kill approximately 32 people every day\cite{NHTSA2020}. In 2022, the NHTSA reported that 32\% of all traffic fatalities involve intoxicated driving\cite{NHTSA2022}. Accidents caused by alcohol have cost the United States economy \$34 billion annually\cite{blincoe2015economic}. Alcohol can impair motor functions and decision-making and lead to risky behaviors such as speeding and running red lights. Alcohol significantly increases the risk of a crash and the severity of it\cite{doi:10.1080/15389588.2013.822491}.

Various countermeasures are employed to combat this issue, including lower legal alcohol limits, public education, and ignition interlock devices (in-vehicle devices that prevent the vehicle from starting if a certain breath alcohol concentration (BrAC) threshold is not met)\cite{doi:10.1080/15389580490465418}. Technological advancements, like continuous intoxication monitoring and real-time driver feedback systems, offer new possibilities for preventing alcohol-related crashes. With growing interest from regulators like EuroNCAP\cite{euroNCAP2023} and the National Transportation Safety Board \cite{NTSB2023, NTSB2022}, the mandatory inclusion of such systems in vehicles is being pushed.

A bipartisan provision in the Infrastructure Investment and Jobs Act requires new national safety standards for passive impaired driving prevention systems in new vehicles\cite{infrastructure2021}. The U.S. government has laid a timeline for the National Highway Traffic Safety Administration (NHTSA) to establish rules requiring passive drunk driving detection systems in new cars. The law gives the NHTSA three years to develop and finalize the regulations for these systems and two to three years for car manufacturers to implement them. This means that drunk driving detection systems could start appearing as standard features in new vehicles as early as 2026 or 2027.

This research aims to assess how drivers feel about integrated monitoring systems and whether drivers feel a sense of security or experience a loss of freedom when they are under surveillance. It is a complex balance, as the presence of monitoring could enhance safety because drivers know that reckless driving is being watched and prevented. But on the other hand, it raises concerns about autonomy, privacy, and how comfortable people feel when every move behind the wheel is recorded. By evaluating how drivers feel about this, we hope to understand the trade-offs drivers are willing to accept in exchange for improved safety.

\section{Related Research}

Real time integrated safety systems in vehicles enhance road safety and reduce accidents caused by impaired or distracted driving. Various studies have explored different approaches to addressing these challenges.

\subsection{Detection Technologies}
The detection of intoxicated driving involves various methods and technologies, each suited to different use-case scenarios, categorized in this study as detection through visual features, biological features, or vehicle movement.

\subsubsection{Detection Through Visual Features}
A method for the detection of intoxication is through machine learning over visual features. Camera-based monitoring utilizes driver-monitoring cameras to analyze eye movements, gaze events, and head movements. The REMoDNaV (Robust Eye Movement Detection and Visualization) algorithm is an algorithm used to extract gaze events such as fixations and saccades (rapid eye movement that moves the center of gaze from one part of the visual field to another part) \cite{10.1145/3544548.3580975}. The data is then processed into time-series segments using a sliding window approach which generates features through statistical aggregation. The system uses logistic regression with Lasso regularization to predict drunk driving at two different BAC thresholds\cite{10.1145/3544548.3580975}. The data is categorized as ``Early Warning" and ``Above Limit" (exceeding the WHO's 0.05 g/dL limit). The model evaluation is based on AUROC. The results are validated through leave-one-subject-out cross-validation. The performance of this model had an overall AUROC of 0.88 $\pm  
0.09$ for ``Above Limit"\cite{10.1145/3544548.3580975}. The system performs similarly across different driving environments with little variation in the mean AUROC. However, the system is more sensitive to moderate alcohol levels. 

Alcohol impairment can be detected through visual features and machine learning, as changes in gaze direction and eyelid opening are indicative of intoxication. Eye-tracking systems process this data to identify patterns such as fixations and saccades, using methods like the Identification by Random Forest (IRF) to detect these events. Glances can be quantified by defining areas of interest, and statistical tools such as Linear Mixed-Effects Regression \cite{Gałecki2013} and Intraclass Correlation Coefficients \cite{Leyland2023} are used to measure variance explained by grouping structures. Analysis shows that with increasing intoxication levels, eye blinks become slower and longer, while glances and fixations become fewer but extend in duration. These changes shift attention to the road at the expense of peripheral areas, reflecting reduced situational awareness. Metrics like the AUROC reveal performance differences in detection setups, with values ranging from 0.73 to 0.81 depending on system configurations. Lower-level psychophysiological measures, such as fixation rates, more accurately capture the driver’s state. However, reliance on eye tracking introduces variability and noise, influenced by hardware configurations \cite{https://doi.org/10.1049/itr2.12520}. 

\subsubsection{Detection Through Breath \& Biological Features}

Breath samples can be used to detect if a driver is impaired. As alcohol enters the bloodstream, it evaporates into the alveolar air in the lung\cite{freudenrich2011breathalyzers}. The ratio between breath alcohol and blood alcohol is typically 2100:1. This makes it possible to estimate BAC by measuring alcohol levels in a driver's breath. This core principle is utilized in three main types of breath alcohol testing devices used in interlock systems: breathalyzers, intoxilyzers, and alcosensors. Each employs different ways of detecting alcohol. Breathalyzers use a chemical reaction to detect alcohol. Potassium dichromate is used because it produces a color change proportional to the alcohol concentration in the breath.\cite{freudenrich2011breathalyzers}. Intoxilyzers use infrared spectroscopy to measure ethanol-specific wavelengths based on molecular vibrations\cite{10.1093/jat/33.2.109}. Alcosensors rely on fuel cells that oxidize alcohol and generate electric currents which can be used to estimate a driver's BAC. The use of infrared spectroscopy offers another advanced sensor technology capable of measuring alcohol levels through breath analysis that adds a layer of precision in detecting intoxication without direct contact. This study used infrared spectroscopy to detect $CO_2$ and alcohol concentrations\cite{doi:10.1080/15389588.2017.1312688}. Different wavelengths of 9.5 µm for alcohol and 4.26 µm for CO2 were used and integrated into a steering column. It was operated at a 5Hz repetition rate. The system was tested using bench and in-vehicle experiments while using human subjects and controlled gas generators. During bench testing, CO2 and alcohol were measured at different distances from the source. It shows that dilution effects occurred as the distance increased. The dilution factor was calculated and derived from the BrAC. Noise limits (0.002 mg/L RMS) set an upper threshold for signal resolution at high dilution levels. 

Infrared spectroscopy enables contactless detection of alcohol levels, and genetic algorithms can enhance detection accuracy by refining signal analysis in resource-constrained environments. A 2021 study used seven alcohol sensors installed in a vehicle to measure alcohol vapor presence\cite{s21227752}. Data was collected using MQ3 sensors, standardized to address variability in sensitivity and signal degradation. Machine learning models were trained to detect patterns, with a genetic algorithm optimizing feature selection for accurate classification. The final model, constructed with a support vector machine (SVM) classifier, achieved a high accuracy score of 0.99, with a 95\% confidence interval ranging from 0.98 to 1.0. Controlled experiments validated the results, employing alcohol samples and various sensor placements. Cross-validation strategies ensured robustness, preventing overfitting and fine-tuning model accuracy. These methods effectively combine sensor data and machine learning to detect impaired driving through breath and biological features.

\subsubsection{Detection Through Vehicle Motion}
DetecDUI is a detection approach through steering wheel \& vehicle motion. It analyzes movement through psychomotor coordination using WiFi signals and an IMU sensor attached to the steering wheel\cite{9626572}. The system's architecture has two main modules: signal processing and fusion. It cleans and processes data while an impaired driving analyzer extracts features to estimate BAC and determine if the driver is intoxicated or not. The system filters out noise, eliminates multipath reflection, uses principal component analysis, and uses an adaptive variational mode decomposition (a method used to separate mixed signals into multiple intrinsic modes) to separate vital signs from environmental noise\cite{9626572} to detect impaired driving through motion.

\subsection{System Responses}

All of these technologies have a system response once they detect intoxicated driving. One such response is an alcohol ignition interlock, which is a DWI (driving while intoxicated) control device that prevents drivers from starting their vehicle if it detects that the driver is under the influence of alcohol\cite{Marques2009}. The engine will not start unless a breath alcohol concentration (BrAC) less than the level set in the device is detected. This is usually between 0.02 g/dL and 0.04 g/dL. Such measurements are commonly taken by breath alcohol detection devices such as breathalyzers.

Visual-based systems provide continuous feedback by monitoring patterns such as eye movements and hand placement, delivering alerts through audio or visual cues. Biological detection technologies, like infrared sensors, notify users when alcohol is detected through the skin, with notifications also sent to law enforcement if tampering or impairment is identified. As previously mentioned, IMU and gas sensor-based approaches send alerts directly to the driver's smartphone. Hybrid systems integrate these feedback mechanisms by combining various detection methods, such as breath sensors and vehicular motion, to deliver graduated alerts based on the severity of impairment. These systems may first issue a warning to the driver, then progressively restrict vehicle speed or deactivate the ignition if unsafe behavior persists\cite{10.1145/3549206.3549268}.

\subsection{Effectiveness \& Limitations}

Research conducted in New Mexico demonstrated that first-time offenders who installed interlock devices had significantly lower recidivism rates (2.6\% per year) compared to those without interlocks (7.1\% per year). However, once the devices were removed, the recidivism rate for the interlock group increased to 4.9\%\cite{doi:10.1080/15389580701598559}. This narrowed the gap with non-interlock users but was still statistically insignificant. These findings align with a systematic review of the effectiveness of ignition interlocks in reducing alcohol-impaired driving-related crashes. It showed that interlocks consistently reduce re-arrest rates while installed. However, the benefits diminish after the interlocks are removed. The review also found that alcohol-related crashes decrease during interlock use, but long-term deterrence remains limited due to low participation rates and the lack of sustained behavior change once the devices are removed \cite{ELDER2011362}. Both studies emphasize the effectiveness of interlocks as a temporary preventive tool, but they show that the long-term benefits are a challenge.

Alcohol detection technologies face different challenges in their real-world application. Laser spectroscopy effectively captured both alcohol and $CO_2$ signals. However, it showed significant limitations with increasing distances, leading to variability in signal quality and systematic overestimation of BrAC values. Noise limits at high dilution levels constrained its resolution, reducing accuracy\cite{doi:10.1080/15389588.2017.1312688}

The DetectDUI system showcased impressive effectiveness, achieving a drunk driving detection accuracy of 96.6\% and demonstrating reliable BAC predictions with a mean absolute error (MAE) between 0.002\% and 0.005\%\cite{9626572}. However, the system faced limitations, such as potential performance drops influenced by emotional states, like sadness, which slightly impacted the accuracy.
 
Signal acquisition and fusion methods were able to enhance sobriety checkpoints by quickly screening large groups. However, these methods suffered from reduced accuracy (78-83\%) at higher BAC levels, with a corresponding rise in false negatives. This limitation shows the need to use passive sensors as a complementary tool rather than a primary one for law enforcement\cite{doi:10.2105/AJPH.83.4.556}. 

Machine learning models employing alcohol vapor sensors within vehicles also revealed certain constraints. While the SVM classifier achieved near-perfect accuracy through optimized feature selection, the study identified potential variability stemming from differences in vehicle conditions, ventilation, and sensor placement\cite{s21227752}. These factors could affect system performance. This shows the need for future adaptations to adapt to various environments and ensure reliability across many different scenarios.

\begin{table*}
\centering
\caption{Survey Questions on Driver Monitoring Systems}
\resizebox{\textwidth}{!}{ % Resize to fit the width of the text
\begin{tabular}{|c|p{14cm}|} % Add an additional column for question labels
\hline
\textbf{No.} & \textbf{Survey Questions} \\ \hline
$q_1$ & How concerned are you about the risks associated with OTHERS on the roadway driving under the influence of alcohol? \\ \hline
$q_2$ & Considering your own driving decisions, how concerned are you about the risks associated with YOU driving under the influence of alcohol? \\ \hline
$q_3$ & How concerned are you about the risks associated with OTHERS on the roadway driving drowsy? \\ \hline
$q_4$ & Considering your own driving decisions, how concerned are you about the risks associated with YOU driving while drowsy? \\ \hline
$q_5$ & How concerned are you about the risks associated with OTHERS on the roadway driving distracted? \\ \hline
$q_6$ & Considering your typical driving behavior, how concerned are you about the risks associated with YOU driving while distracted? \\ \hline
$q_7$ & Considering your typical use of alcohol, how concerned are you about risks associated with decisions you typically make when deciding whether or not to drive? (For example, if you never consume alcohol or always take a cab after drinking, your response may be "Not concerned at all". Your typical decisions around moderation, time since last drink, and metabolism may be reflected in your answer.) \\ \hline
$q_8$ & Are you aware of technologies that monitor driver behavior to prevent accidents? \\ \hline
$q_9$ & How comfortable would you feel having a system that monitors your eye movements while driving? \\ \hline
$q_{10}$ & How comfortable would you feel with a system that tracks your posture and hand placement while driving? \\ \hline
$q_{11}$ & How comfortable are with having an integrated BrAC level monitoring system that locks the ignition of your car? \\ \hline
$q_{12}$ & How comfortable are with having a passive integrated BrAC level monitoring system that monitors your BrAC level throughout your drive? \\ \hline
$q_{13}$ & How much trust would you place in the accuracy of a system that monitors your eyes, posture, hands, and BrAC? \\ \hline
$q_{14}$ & How likely are you to change your driving behavior based on the alerts or warnings from a driver monitoring system? \\ \hline
$q_{15}$ & In terms of fault tolerance, what rate of "false positive" errors, where a system mistakenly predicts the driver is under the influence of alcohol, would you be comfortable with? \\ \hline
$q_{16}$ & What percentage of instances of driving under impairment of alcohol are you ok with the system missing? \\ \hline
$q_{17}$ & How should OTHER PEOPLE'S vehicles respond when they detect alcohol-impaired driving? \\ \hline
$q_{18}$ & How would YOU like YOUR vehicle to respond when it detects alcohol-impaired driving? \\ \hline
$q_{19}$ & Would you prefer the data collected to be processed locally or through the cloud? \\ \hline
$q_{20}$ & In a driver monitoring system, would you prefer your identity to be anonymized before being processed locally? \\ \hline
$q_{21}$ & In a driver monitoring system, would you prefer your identity to be anonymized before being processed in the cloud? \\ \hline
$q_{22}$ & Would the presence of a monitoring system that tracks eyes, posture, hands, and BrAC influence your decision to purchase a vehicle? \\ \hline
$q_{23}$ & Would you be willing to pay extra for a vehicle equipped with such advanced monitoring technologies? \\ \hline
$q_{24}$ & How appealing do you find the idea of having a comprehensive driver monitoring system in your vehicle? \\ \hline
$q_{25}$ & What do you think about using technology to monitor drivers' eyes, posture, hand placement, and BrAC (breath-alcohol concentration) to prevent accidents? \\ \hline
$q_{26}$ & Do you have any other thoughts or concerns about using driver monitoring systems to enhance road safety? \\ \hline
\end{tabular}
\label{questiontable}

}
\end{table*}

\subsection{Adoption of Technologies}

The use of interlocks has risen sharply since 2006. Many organizations, such as Mother Against Drunk Driving, call for interlock systems to be a central feature of vehicles. They call for mandatory ignition locks in all states and provinces\cite{MADD_2022}. 

The Infrastructure Investment and Jobs Act mandates the National Highway Traffic Safety Administration (NHTSA) to establish a Federal Motor Vehicle Safety Standard requiring new passenger vehicles to be equipped with advanced drunk and impaired driving prevention technology. This technology is intended to passively monitor a driver's performance to accurately identify impairment and prevent or limit vehicle operation if impairment is detected\cite{infrastructure2021}

\subsection{Cost}

Many pilot programs have begun implementing interlock systems into vehicles. Within the European Union, countries such as France and Sweden have started to adopt interlock systems by implementing them into vehicles for alcohol offenders. The cost is about \$2.20 USD per day, along with a fee of \$125 USD to install such interlock systems\cite{Marques2009}. 

After a search of the existing literature on implementations of machine learning and sensor-based systems, there do not seem to be conclusive studies on the costs of these different technologies, which could be related to their newness.

\section{Method}

To understand driver perspectives on monitoring systems, we administered a survey to U.S. residents. Participants were recruited through flyers posted at research labs and outreach via local networks, including individuals in tech and education. We collected responses to the questions shown in Table \ref{questiontable}—covering awareness of risks, familiarity with existing monitoring technologies, level of comfort and trust with technology, privacy concerns, vehicle purchase decisions, and an open opportunity to share any additional thoughts. Following IRB protocol, we also collected informed consent and demographic information from every participant.

Demographic information included gender, age, ethnicity, education, marital status, employment, and income. These variables were essential to compare responses across different demographics. Collecting such data allowed the assessment of how perspectives on monitoring systems varied based on age, education, or car type. Asking about the level of automation in the driver’s current vehicle gave context to how familiar a driver was with such technology and provided insight into its effect on their viewpoint of monitoring systems.

Driver awareness questions assessed respondents’ awareness of the risks associated with driving while impaired. It provided insights into the driver’s perception of risk regarding impaired, distracted, or drowsy driving. It differentiated how worried they were about themselves from how worried they were about others. The goal was to understand how self-aware the driver was and how they perceived the danger posed by others. The question’s emphasis on both personal and external concerns around alcohol, drowsiness, and distractions revealed the drivers' priorities. It could guide what aspects of safety systems should be emphasized—whether to target personal accountability or mitigate risks caused by others.

Respondents were asked how familiar they were with monitoring systems to assess how informed they were about such systems. It also addressed how comfortable they felt using systems that monitored their eyes, posture, hand movements, and alcohol levels. The questions aimed to gauge how they felt about surveillance technology and the intrusion of their privacy. The section touched on trust and whether the drivers thought the systems would work reliably. These questions showed the trade-off between safety and the potential for system errors, such as false positives or false negatives. This helped to evaluate whether people would accept or reject monitoring systems.

Drivers’ privacy concerns surrounding monitoring systems were needed to evaluate how much privacy impacted respondent perceptions. It asked about anonymity when data was collected and processed, addressing whether privacy was a primary concern when users adopted such systems.

Assessing how much weight drivers gave to monitoring systems in their decision to buy vehicles allowed us to get to the commercial side of monitoring technologies: whether people believed these systems were beneficial or harmful enough to influence their buying decisions.

\subsection{Results}

%score

% \begin{table*}
% \centering
% \caption{Demographic Information Summary}
% \adjustbox{width=\textwidth}{
% \begin{tabular}{|l|l|l|l|l|l|l|l|l|}
% \hline
% Gender & 60.5\% (Male) & 39.5\% (Female) &  &  &  &  &  &  \\ \hline
% Age & 31.3\% (35-44) & 21.7\% (45-54) & 20.9\% (24-34) & 13.9\% (18-24) & 8.7\% (55-64) & 3.5\% (65+) &  &  \\ \hline
% Ethnicity & 58.4\% (Asian) & 32.7\% (White/Caucasian) & 3.5\% (African-American) & 3.5\% (Two or more) & 1.8\% (Hispanic) &  &  &  \\ \hline
% Education & 45.2\% (Master's degree) & 36.5\% (Bachelor's degree) & 7.0\% (Doctoral degree) & 7.0\% (Some college) & 2.6\% (Technical certification) & 0.9\% (Associate degree) & 0.9\% (High School) &  \\ \hline
% Marital Status & 67.5\% (Married) & 29.8\% (Single) & 1.8\% (Divorced) & 0.9\% (Separated) &  &  &  &  \\ \hline
% Employment & 79.8\% (Full-time) & 7.0\% (Unemployed) & 6.1\% (Part-time) & 2.6\% (Contract or temporary) & 1.8\% (Retired) & 0.9\% (Studying in college) & 0.9\% (Home Maker) & 0.9\% (Student) \\ \hline
% Annual household income & 33.6\% (Prefer not to say) & 31.0\% (\$150,000+) & 15.9\% (\$120,000-\$149,999) & 7.1\% (\$60,000-\$89,999) & 5.3\% (\$30,000-\$59,999) & 3.5\% (\$90,000-\$119,999) & 3.5\% (\$0-\$29,999) &  \\ \hline
% Reported Automation Level & 54.4\% (Level 1) & 16.7\% (Level 0) & 16.7\% (Level 2) & 7.0\% (Level 3) & 2.6\% (Level 4) & 2.6\% (No Vehicle) &  &  \\ \hline
% \end{tabular}
% }
% \label{tab:demographics}
% \end{table*}

\begin{table*}
\centering
\caption{Demographic Information Summary}
\adjustbox{width=\textwidth}{
\begin{tabular}{|l|l|l|l|l|l|l|l|}
\hline
\textbf{Gender} & \textbf{Age} & \textbf{Ethnicity} & \textbf{Education} & \textbf{Marital Status} & \textbf{Employment} & \textbf{Household Income} & \textbf{Automation Level} \\ \hline
  Male 60.5\% & 18-24 13.9\% & Asian 58.4\% & High School 0.9\% & Married 67.5\% & Full-time 79.8\% & \$0-\$29,999 3.5\%& Level 0 16.7\%\\ \hline
  Female 39.5\%  & 24-34 20.9\% &White/Caucasian 32.7\%  &Some college 7.0\%  & Single 29.8\% & Part-time 6.1\% & \$30,000-\$59,999 5.3\%  & Level 1 54.4\% \\ \hline
  & 35-44 31.3\% & African-American 3.5\% & Technical Certificate 2.6\% & Divorced 1.8\% & Unemployed 7.0\% & \$60,000-\$89,999 7.1\% & Level 2 16.7\%  \\ \hline
   & 45-54 21.7\% & Hispanic 1.8\% & Associate degree 0.9\%  & Separated 0.9\% & Contract 2.6\% & \$90,000-\$119,999 3.5\% & Level 3 7.0\% \\ \hline
   & 55-64 8.7\% & Two or more 3.5\% & Bachelor's degree 36.5\%&  & Retired 1.8\% &\$120,000-\$149,999 15.9\% & Level 4 2.6\% \\ \hline
   & 65+ 3.5\% &  & Master's degree 45.2\%  &  & Student 1.8\% & \$150,000+ 31.0\% & No Vehicle 2.6\% \\ \hline
   &  &  & Doctoral degree 7.0\%  &  & Home Maker 0.9\% & Prefer not to say 33.6\%  &  \\ \hline
   % &  &  &  &  & Student 0.9\% &  &  \\ \hline
   % &  &  &  &  &  &  &  \\ \hline
\end{tabular}
}
\label{tab:demographics}
\end{table*}

% \begin{table*}
% \centering
% \caption{Demographic Information Summary (Reformatted)}
% \adjustbox{width=\textwidth}{
% \begin{tabular}{|l|l|l|l|l|l|l|l|}
% \hline
% Gender & Male 60.5\% & Female 39.5\% &  &  &  &  &  \\ \hline
% Age & 18-24 13.9\% & 24-34 20.9\% & 35-44 31.3\% & 45-54 21.7\% & 55-64 8.7\% & 65+ 3.5\% &  \\ \hline
% Ethnicity & Asian 58.4\% & White/Caucasian 32.7\% & African-American 3.5\% & Two or more 3.5\% & Hispanic 1.8\% &  &  \\ \hline
% Education & High School 0.9\% & Associate degree 0.9\% & Technical certification 2.6\% & Some college 7.0\% & Bachelor's degree 36.5\% & Master's degree 45.2\% & Doctoral degree 7.0\% \\ \hline
% Marital Status & Married 67.5\% & Single 29.8\% & Divorced 1.8\% & Separated 0.9\% &  & &  \\ \hline
% Employment & Full-time 79.8\% & Unemployed 7.0\% & Part-time 6.1\% & Contract or temporary 2.6\% & Retired 1.8\% & Studying in college 0.9\% & Home Maker 0.9\% & Student 0.9\% \\ \hline
% Annual Household Income & \$0-\$29,999 3.5\% & \$30,000-\$59,999 5.3\% & \$60,000-\$89,999 7.1\% & \$90,000-\$119,999 3.5\% & \$120,000-\$149,999 15.9\% & \$150,000+ 31.0\% & Prefer not to say 33.6\% \\ \hline
% Reported Automation Level & Level 0 16.7\% & Level 1 54.4\% & Level 2 16.7\% & Level 3 7.0\% & Level 4 2.6\% & No Vehicle 2.6\% &  \\ \hline
% \end{tabular}
% }
% \label{tab:demographics-reformatted}
% \end{table*}

The survey garnered 115 respondents. The demographic information from the survey can be found in Table \ref{tab:demographics}.

The majority expressed mixed feelings and varying comfort levels with driver monitoring systems. Driver concern about others' driving was heavily skewed towards being extremely concerned, with 55.65\% of respondents indicating high levels of concern. For concern about their own driving behavior, 25.22\% expressed significant concern. 30.26\% of the respondents were aware of existing monitoring systems. 17.39\% reported reading extensively into these technologies.

The score represents a value out of 5 on a scale of 1-5. The trust in the accuracy of these systems had an average score of 3.03. 34.58\% rated their trust at 3 out of 5. Drivers' perception of the likelihood of changing driving behavior after receiving warnings averaged 3.75. 52.08\% indicated a positive response to warnings.

Comfort levels with different monitoring systems varied. Systems that monitor eye movements had a mean score of 3.24 with a standard deviation of 1.23. 48.96\% felt somewhat comfortable. Posture and hand placement tracking systems had an average score of 3.17 and a standard deviation of 1.30. 41.67\% reported moderate comfort. Integrated BrAC systems that lock the ignition based on the driver actively blowing on the breathalyzer before driving had a mean comfort rating of 3.72 and a standard deviation of 1.32. 54.39\% responded positively. Passive BrAC monitoring systems that track BrAC levels passively without direct user interference throughout the drive had a mean score of 3.66 with a standard deviation of 1.22. 50.87\% indicated comfort. Drivers showed an average willingness-to-pay score of 2.57, indicating they were generally unlikely to pay extra for these systems.

\begin{figure}
    \centering
    \includegraphics[width=0.5\textwidth]{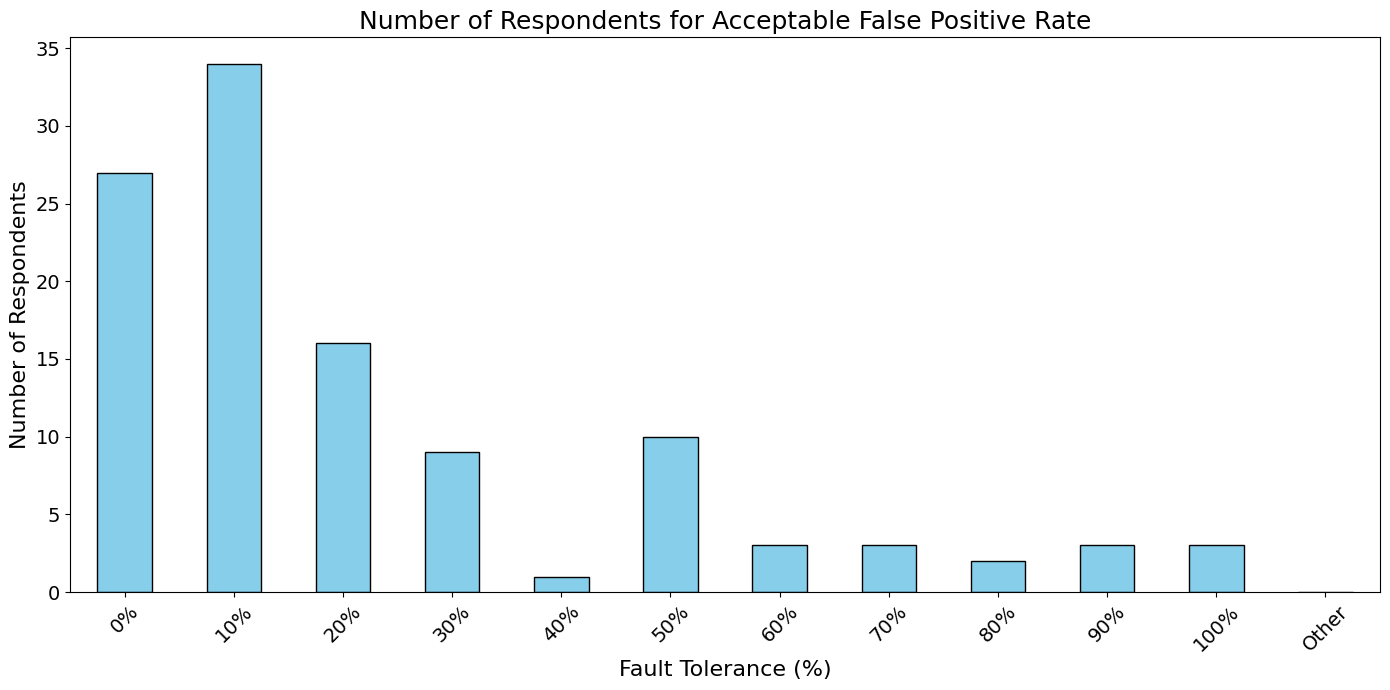}
    \caption{This graph compares the number of respondents that chose each false positive percentage as acceptable. Participants tended to favor lower percentages of false positives.}
    \label{fig:false_positives}
\end{figure}

\begin{figure}
    \centering
    \includegraphics[width=0.5\textwidth]{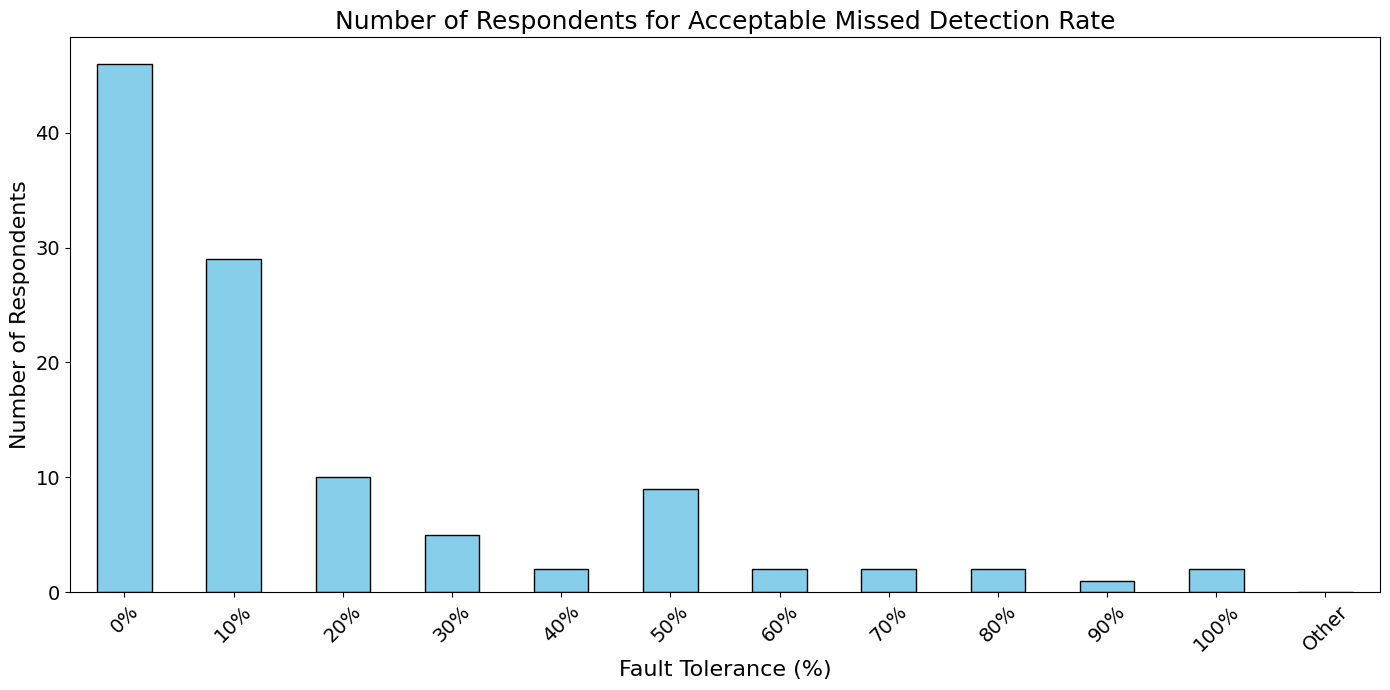}
    \caption{This graph compares the number of respondents that chose each acceptable missed detection percentage. Participants tended to favor lower percentages of missed detections.}
    \label{fig:missed_detections}
\end{figure}

The average acceptance of comprehensive driver monitoring systems was 3.23. 45.83\% felt moderately accepting. Privacy concerns were expressed by 76.52\% who preferred identity anonymization during local data processing. 83.33\% preferred anonymization before cloud processing. For data processing preferences, 61.74\% chose local processing. 29.57\% were indifferent. 8.70\% preferred cloud-based processing. As seen in Figure \ref{fig:false_positives} for acceptable ``false positive" error rates, 29.82\% were comfortable with a 10\% error rate. 23.68\% expected a 0\% error rate. For acceptable impairment detection misses in Figure \ref{fig:missed_detections}, 40.71\% chose 0\% and 25.66\% accepted up to 10\%.

Regarding how other people's vehicles should respond to alcohol-impaired driving detection, 29.48\% preferred visual warnings, 28.69\% chose audio warnings, 25.10\% selected the car pulling over, 13.55\% favored contacting authorities. For responses about their own vehicle's response, 30.74\% preferred audio warnings, 28.69\% chose visual warnings, 27.87\% wanted the car to pull over, 9.02\% selected contacting authorities.

\begin{figure}
    \centering
    \includegraphics[width=0.5\textwidth]{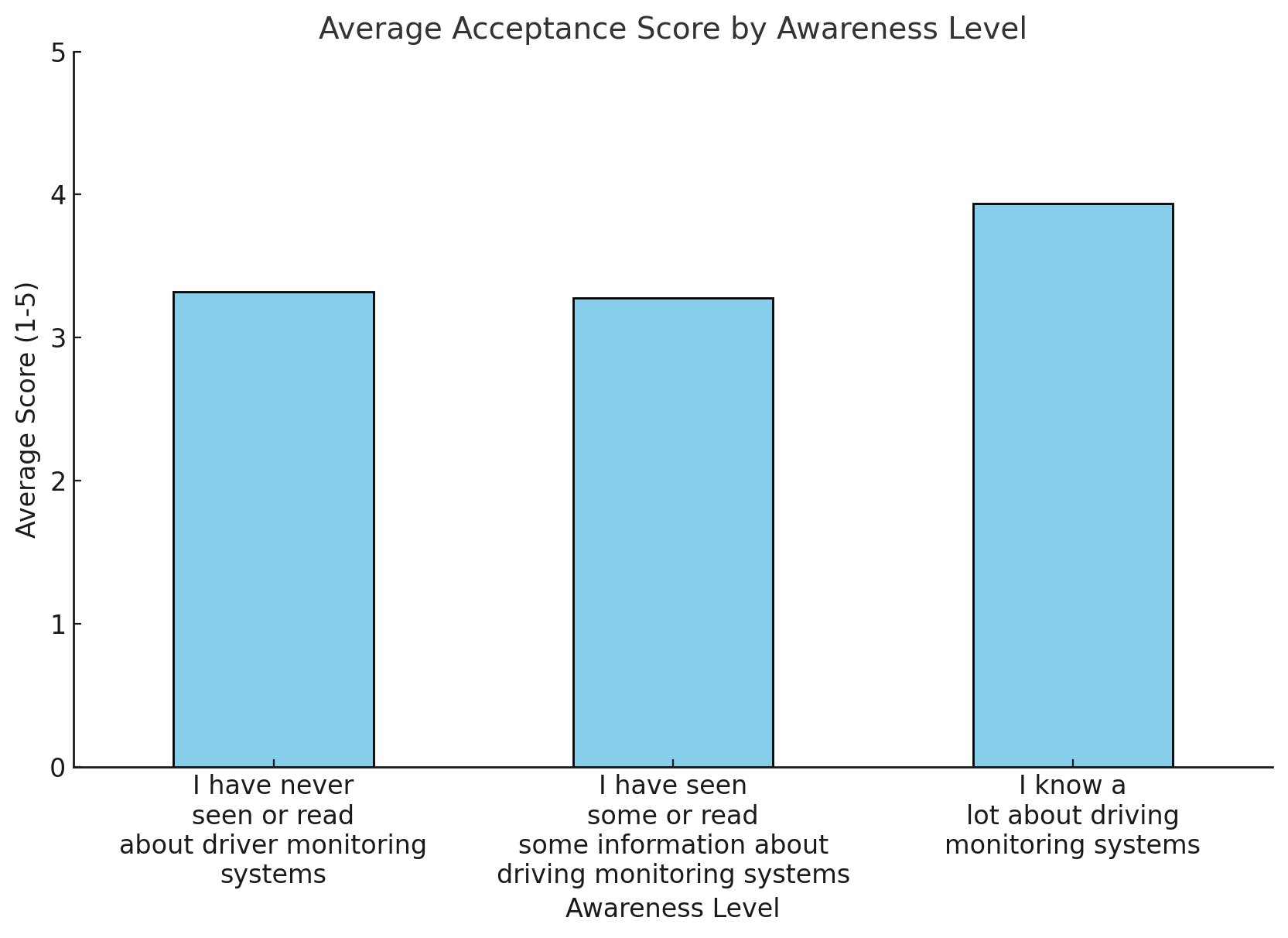}
    \caption{This graph compares the average points for the Acceptance Score of each individual depending on the level of awareness. Those who have read or seen a lot of information on monitoring systems show a higher level of acceptance than those who have read or seen either some or no information on such systems.}
    \label{fig:awareness_acceptance}
\end{figure}
\begin{figure}
    \centering
    \includegraphics[width=0.5\textwidth]{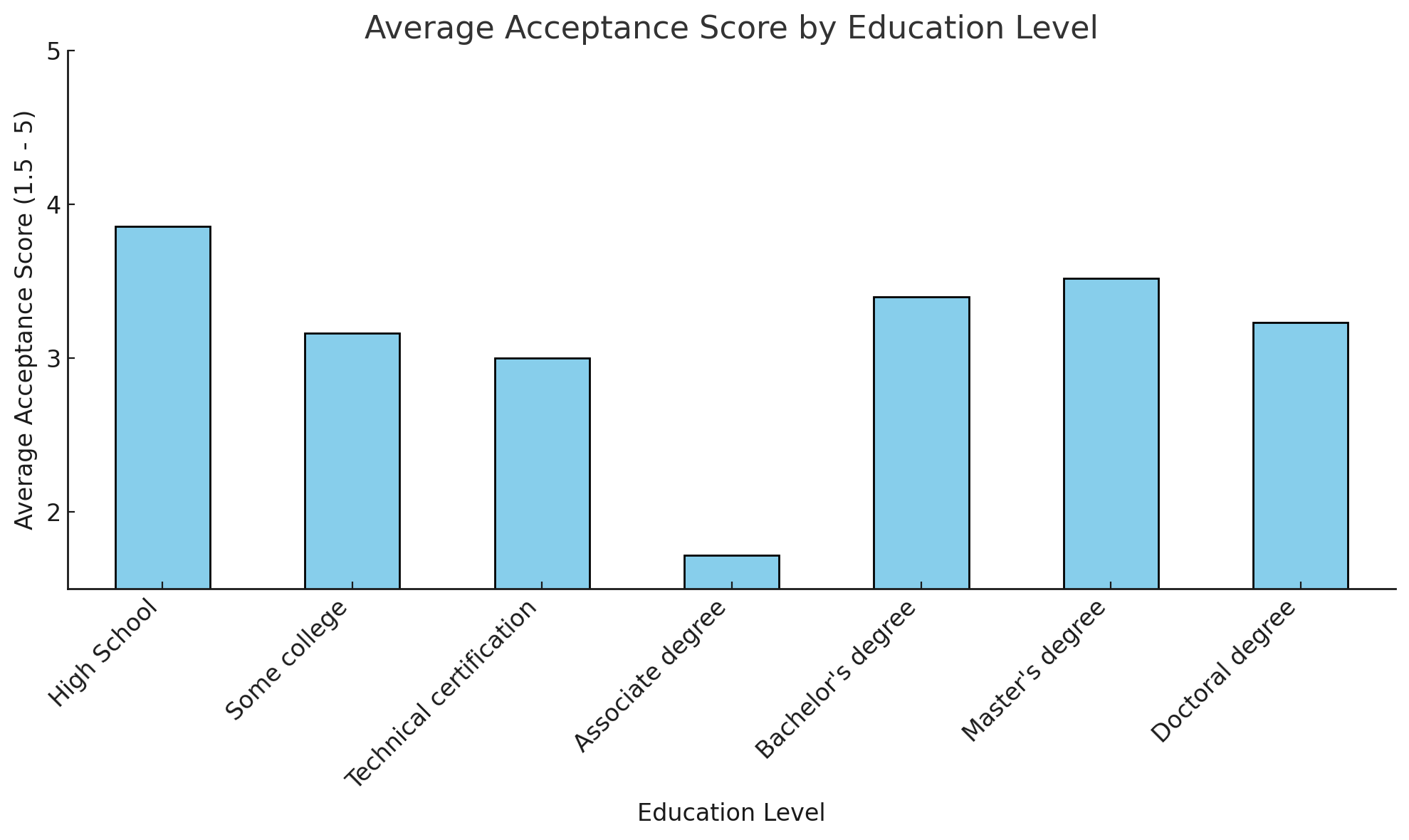}
    \caption{This graph compares the average acceptance score with education level; we observe no strong trend in the data.}
    \label{fig:education}
\end{figure}
\begin{figure}
    \centering
    \includegraphics[width=0.5\textwidth]{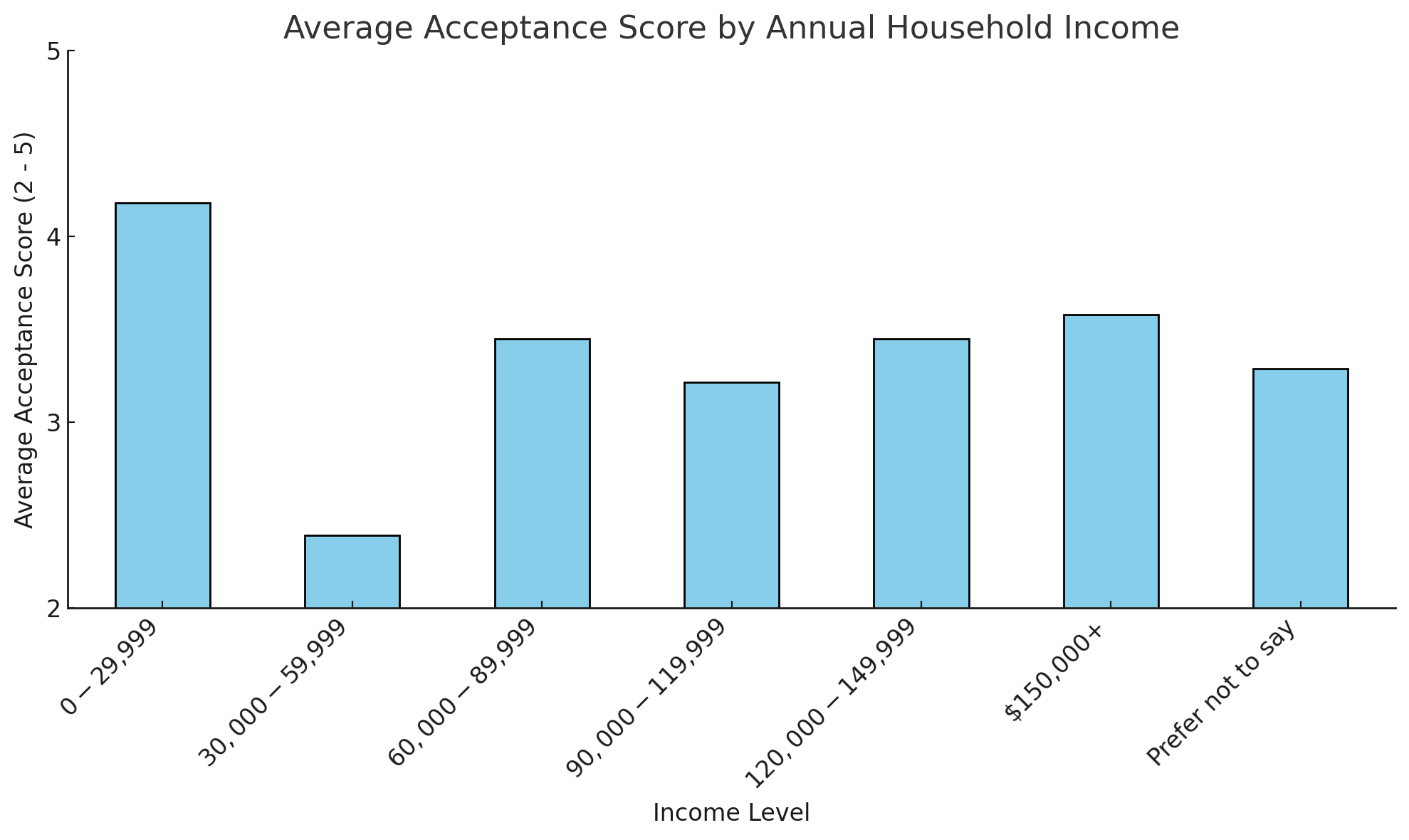}
    \caption{This graph compares the average acceptance score with income level; we observe no strong trend in the data.}
    \label{fig:income}
\end{figure}
\begin{figure}
    \centering
    \includegraphics[width=0.5\textwidth]{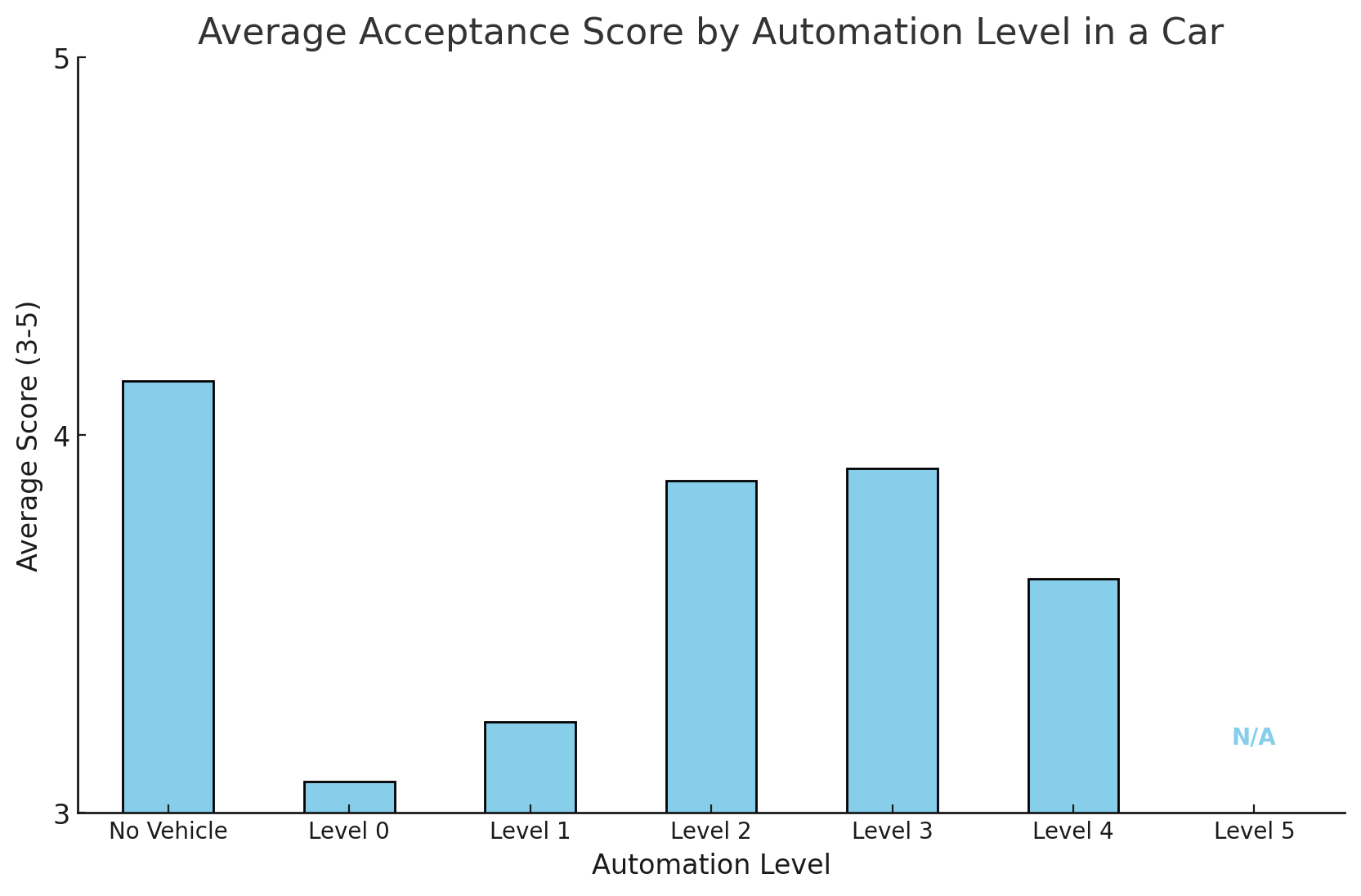}
    \caption{This graph compares the average acceptance score with the automation level of a vehicle. We note that the reported automation level values are perceived automation levels by the respondent, which may not accurately reflect their actual level of vehicle automation. This is discussed in Section IV.}
    \label{fig:automation}
\end{figure}
\begin{figure}
    \centering
    \includegraphics[width=0.5\textwidth]{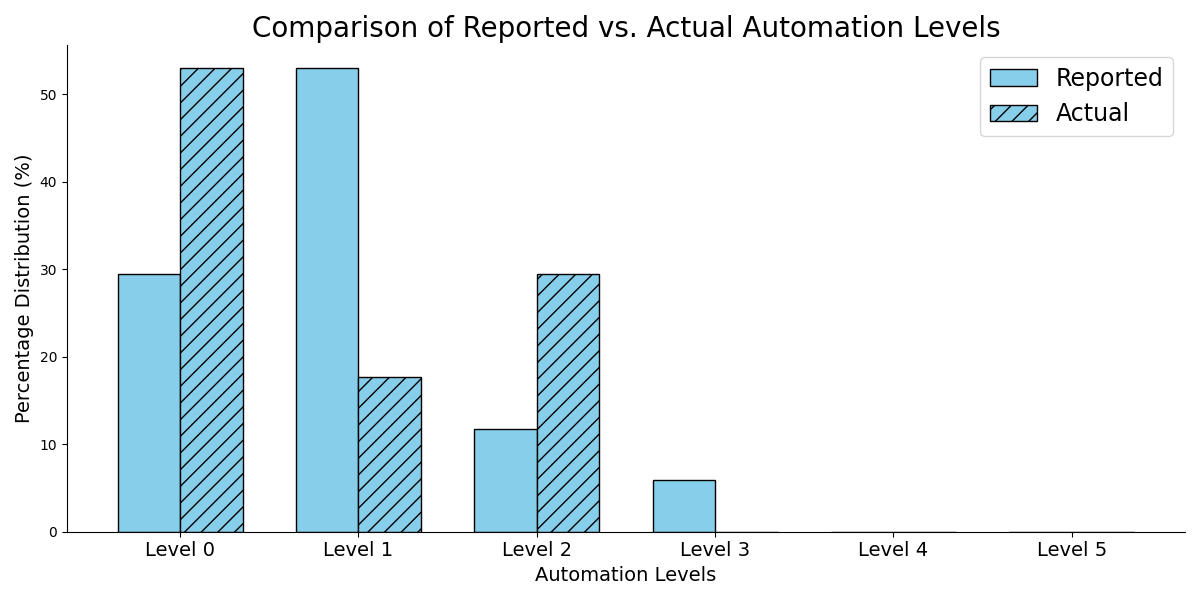}
    \caption{This graph compares the reported automation level with the actual automation level of the vehicles reported. Respondents tended to overestimate the automation level of their vehicle.}
    \label{fig:automation_double}
\end{figure}
\begin{figure}
    \centering
    \includegraphics[width=0.5\textwidth]{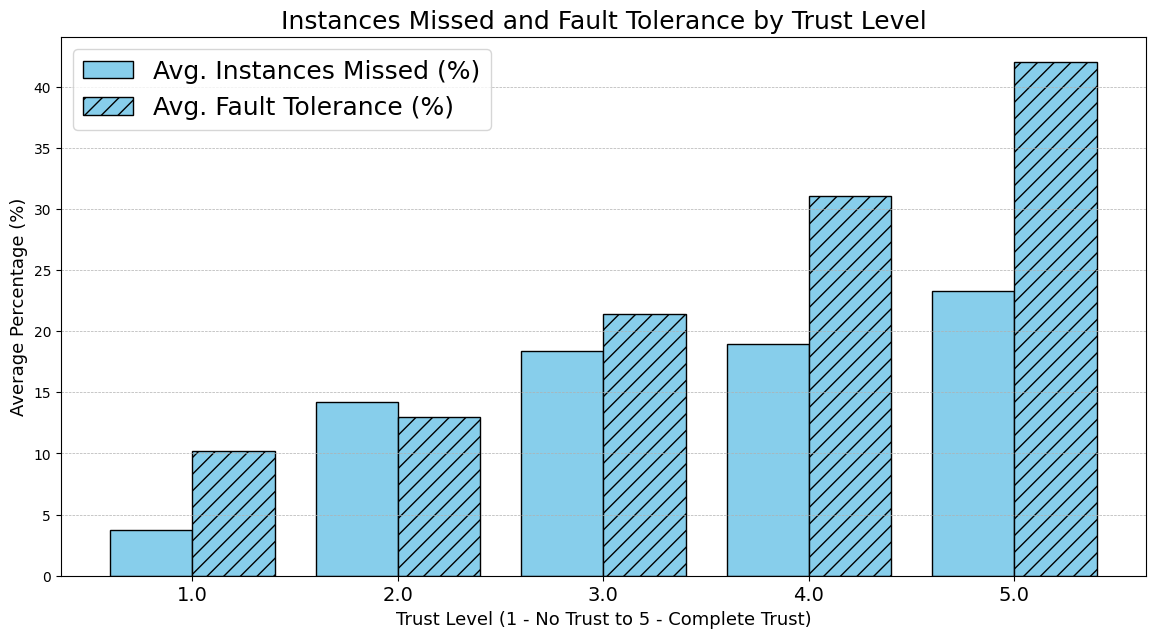}
    \caption{This graph compares the average percentage of false positives \& the average percentage of missed instances of impaired driving depending on the level of trust given to such systems. As the driver's trust in monitoring systems increases, they tend to accept higher false positives. They also tend to accept higher percentages of missed instances of impaired driving.}
    \label{fig:double}
\end{figure}
\begin{figure}
    \centering
    \includegraphics[width=0.5\textwidth]{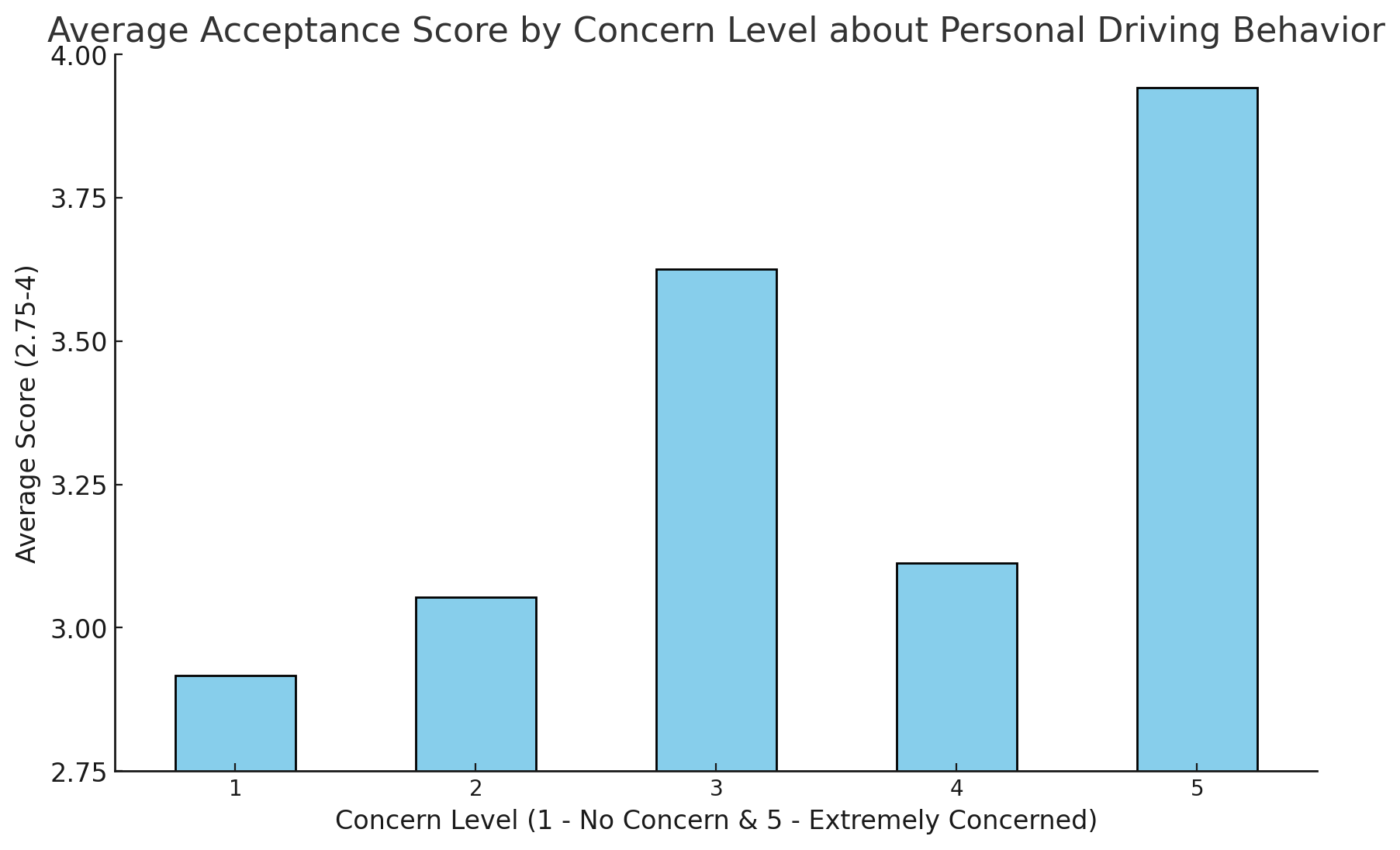}
    \caption{This graph compares the average points for the Acceptance Score of each individual depending on concern about personal driving behavior. A general trend is seen in the acceptance scores: as personal concerns grow so does the general acceptance of driver monitoring systems.}
    \label{fig:personal_acceptance}
\end{figure}

\begin{figure}
    \centering
    \includegraphics[width=0.5\textwidth]{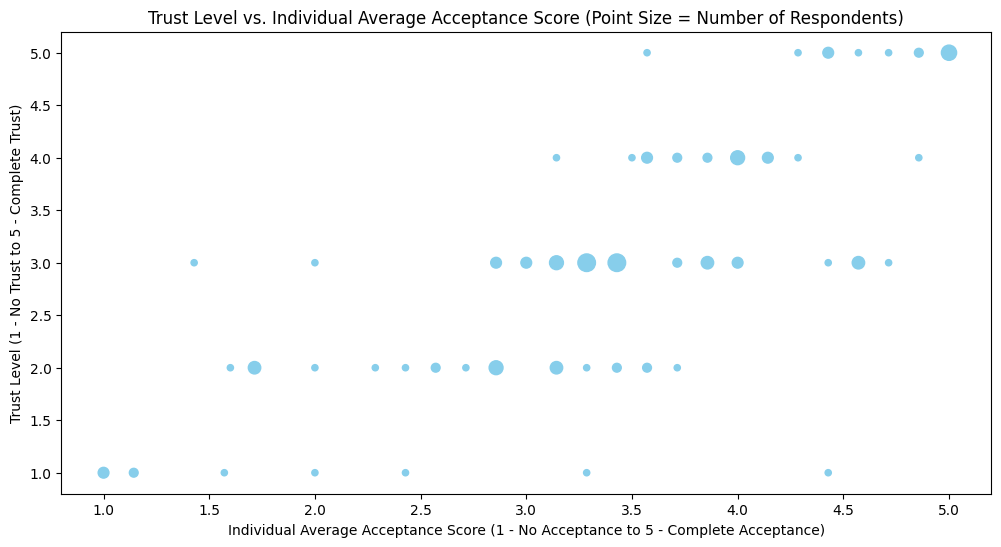}
    \caption{This graph plots each trust level depending on each acceptance score for each individual. The larger the radius of the point, the more respondents that had the same acceptance and trust scores. The trust question ($q_{13}$) was removed from the calculation of the acceptance score to prevent correlating variables from affecting each other. Generally, as driver acceptance of monitoring systems goes up, the amount of trust placed in such systems goes up.}
    \label{fig:trust_acceptance}
\end{figure}

\section{Discussion}

%comment 

Taking into account the above ideas, we define a measure of acceptance of driver monitoring systems called the Individual Acceptance Score (IAS), which takes into account particular survey question responses and ranges in value from 1-5 (1 - No Acceptance \& 5 - Complete Acceptance). To compare different categorical groups of survey subjects, we computed the Average Acceptance Score. The following equations define the two measures: 

\begin{equation} \text{IAS}_i = \frac{q_{9} + q_{10} + q_{11} + q_{12} + q_{13} + q_{14} + q_{24}}{7} \end{equation}
\begin{equation}
\text{AAS}(c) = \frac{1}{n_c} \sum_{i=1}^{n_c} \text{IAS}_i
\end{equation}

AAS(\(c \)) represents the Average Acceptance Score as a function of a particular class of respondents, where \(n \) is the total number of respondents within class $c$.

The results of this survey show public perception towards driver monitoring systems. Respondents who had a greater awareness of such systems were more likely to accept and trust these technologies, as seen in Figure \ref{fig:awareness_acceptance}. This suggests that public exposure and educational implementation can enhance trust and openness to these systems. The more informed a driver is on monitoring systems, the more likely they are to trust it. Interestingly, Figure \ref{fig:education} and Figure \ref{fig:income} show that education level and income level did not have a significant impact on acceptance score. Hence, a certain education level is not needed to accept monitoring systems. Nor is acceptance restricted to those who can afford such systems. Drivers just have to be informed of such systems. 

People preferred noninvasive systems like eye-tracking over invasive ones such as BrAC ignition interlocks, highlighting a value for personal autonomy and freedom.

Figure \ref{fig:automation} shows that the automation level of each driver's vehicle did not significantly impact the acceptance score. This shows that a level of familiarity is not needed to implement such systems. Instead, drivers being informed about monitoring technologies makes them more willing to accept these technologies. However, it is important to note that the reported automation level is based on driver perception. Figure \ref{fig:automation_double} shows an intriguing trend where drivers tended to overestimate their vehicle's actual automation levels. The actual automation level was determined by referencing the driver's manual of each respondent's car with the framework established by the Society of Automotive Engineers (SAE)\cite{sae_j3016}.

Trust significantly influenced participants’ willingness to change their driving behavior based on system feedback. Those who trusted the accuracy of such systems indicated a higher likelihood of changing their driving in response to alerts and warnings from monitoring tools. Figure \ref{fig:double} shows that as trust levels rise, so does the average acceptable false positives \& missed instances. This means that as trust is gained, manufacturers will have looser constraints on the accuracy of monitoring systems. This shows that manufacturers should prioritize transparency and reliability in these systems to improve trust and overall acceptance of the systems. 

Privacy concerns were identified as a significant barrier to acceptance. Most respondents preferred local data processing over other cloud-based solutions. This strong preference for local data handling shows the desire for more control over their personal information and privacy. They also preferred that their identity be anonymized regardless of the type of processing used. Data privacy is a critical issue that regulators and manufacturers need to consider to gain public trust and adoption of these technologies. 

Figure \ref{fig:personal_acceptance} shows a surprising trend where drivers were more willing to accept monitoring systems depending on their driving behavior. Drivers want to use these systems more if they are concerned about their personal driving behavior. This shows that respondents are willing to hold themselves accountable while using these systems.

The tolerance for system errors was relatively low. Respondents were only comfortable with minimal false positives. This highlights the need for more precision and reliability within these systems. Participants favored high accuracy when asked about the acceptable levels of missed detections. However, they become more open to more errors from these systems if they are more comfortable and trusting of them, as seen in Figure \ref{fig:trust_acceptance}. This shows respondent trust can be elevated if they are more accepting of monitoring systems, hence the rate of errors from these systems can be more tolerable. Therefore, systems must gain public trust and acceptance to allow for looser accuracy constraints. The results show that there is a foundational level of support for driver monitoring systems as acceptance levels have been in the 3-4 range out of 5. Drivers are willing to try and begin to implement these systems, but their concerns have to be addressed. The path to widespread public adoption is to address key concerns such as system reliability, privacy, and balance between safety benefits and personal autonomy. Steps can be taken towards these goals through data protection measures and the combination of different technologies to make a more accurate system. Informing and promoting such systems to the public can increase the acceptance and integration of these technologies into everyday vehicle use.

\section{Conclusion}

With the emerging implementation and regulation of integrated drive safety monitoring systems, it is imperative to understand the public's view on such systems and how they can benefit future decision-making. Different technologies are being used to develop diverse monitoring technologies with different use cases and responses to impaired drivers. To evaluate what drivers think of such new technologies, this study implemented a survey that assessed demographics, awareness, comfort, trust, and privacy concerns on monitoring systems. The results indicate that higher familiarity with these technologies correlated with greater acceptance, particularly for non-intrusive systems like eye and posture monitoring, compared to more restrictive measures such as BrAC monitoring. Privacy was a primary concern, and most people favored local data processing and data anonymization. High accuracy and reliability are also needed within these systems to increase driver acceptance. Public trust and willingness to adapt driving behavior based on system alerts were positively correlated to confidence in system accuracy. These concerns can be addressed through transparent data practices, privacy safeguards, and public education to, in turn, bolster trust and acceptance. These strategies can help integrate advanced driver monitoring systems effectively and foster safer driving behaviors that contribute to better road safety.

\bibliographystyle{ieeetr}
\bibliography{refs}

\end{document}